\documentclass[copyright]{eptcs}

\usepackage{iftex}
\usepackage{listings}

\ifpdf
  \usepackage{underscore}         
  \usepackage[T1]{fontenc}        
\else
  \usepackage{breakurl}           
\fi

\usepackage{xurl}

\title{VSS Challenge Problem:\\
Verifying the Correctness of AllReduce Algorithms in the MPICH Implementation of MPI}
\author{Paul D. Hovland
\institute{Mathematics and Computer Science Division\\ Argonne National Laboratory\\ Lemont, IL, USA }
\email{hovland@anl.gov}
}
\def\titlerunning{Verifying the Correctness of AllReduce Algorithms}
\def\authorrunning{P.D. Hovland}
\def\copyrightholders{UChicago Argonne, LLC,\\ Operator of Argonne National Laboratory}
\begin{document}
\maketitle

\begin{abstract}
We describe a challenge problem for verification based on the MPICH implementation of MPI. The MPICH implementation includes several algorithms for allreduce, all of which should be functionally equivalent to reduce followed by broadcast. We created standalone versions of three algorithms and verified two of them using CIVL.
\end{abstract}

\section{Introduction}

The message passing interface (MPI) standard includes several forms of collective communication, including broadcast, allgather, reduce, and allreduce~\cite{10.5555/552013}. Implementations of MPI such as MPICH and OpenMPI often include multiple algorithms for each collective communication operation, with the particular algorithm selected at runtime based on hardware characteristics such as network topology and latencies and application characteristics such as process count and array lengths~\cite{thakur2005optimization}. All of the algorithms for a given operation should be functionally equivalent; the only differences should be in performance and, perhaps, restrictions on the domain of applicability (e.g., only power-of-two process counts or only for builtin (not user-defined) datatypes). Consequently, the correctness of the implementations of the various algorithms can be confirmed by verifying their equivalence to one another and to the behavior specified in the MPI standard.

As a concrete example, we consider the implementation of \texttt{MPI\_Allreduce} in the MPICH library.  This function performs a global reduction operation such as product or sum and distributes the result to all processes. It can be viewed as equivalent to \texttt{MPI\_Reduce} followed by \texttt{MPI\_Broadcast}. MPICH version 4.3.0 implements 7 different algorithms for (blocking) allreduce (not including special cases for intercommunicator reductions and shared memory parallelism): recursive doubling, recursive exchange, recursive multiplying, reducescatter-allgather, k-way reducescatter-allgather, ring exchange, and tree exchange. In addition, MPICH supports calling one of the nonblocking algorithms and waiting on the result. Some of these algorithms are described in~\cite{thakur2005optimization}.

The MPICH implementations of allreduce algorithms typically rely on internal communication routines such as \texttt{MPIC\_Send} and \texttt{MPIR\_Reduce\_local} as well as MPICH-specific data structures. To facilitate verification of individual algorithms, we created a header file that replaces MPICH's \texttt{mpiimpl.h} and redefines MPICH subroutines and macros using standard MPI functions. This enables one to, for example, compile the file \texttt{allreduce\_intra\_recursive\_doubling.c} with no modifictions and without any additional dependencies and use \texttt{MPIR\_Allreduce\_intra\_recursive\_doubling} as a replacement for \texttt{MPI\_Allreduce}. A limitation of this approach is that we assume that the reduction operator is a builtin reduction operator (maximum, sum, product, etc.) and the datatype is a builtin datatype (integer, float, double, etc.). It may be possible to remove the first restriction by making fewer assumptions about commutativity but the latter assumption will be difficult to overcome, in part because one would need to expose or replicate much of the complex mechanisms inside MPICH to deal with derived datatypes.

Currently, our challenge problem\footnote{Available at \url{https://github.com/pmodels/civl-verification}} consists of the following files:
\begin{description}
    \item[\texttt{allreduce\_intra\_recursive\_doubling.c}:] verbatim copy of the MPICH v4.3.0 file, also available at \url{https://github.com/pmodels/mpich/blob/main/src/mpi/coll/allreduce/allreduce\_intra\_recursive\_doubling.c} and included with light editing as Appendix~\ref{app:allreduce}.
     \item[\texttt{allreduce\_intra\_recursive\_multiplying.c}:] slightly modified version of the file from MPICH v.4.3.0 (modifications include replacing direct access to fields within MPICH internal data structures with accessor macros and allocating memory for nonblocking communication requests).
     \item[\texttt{allreduce\_intra\_reduce\_scatter\_allgather.c}:] slightly modified version of the file from MPICH v.4.3.0 (replace direct access to communicator internals and alloc/free temporary data structures).
    \item[\texttt{mpiimpl.h}:] wrapper functions and macros to replace MPICH-specific functions and data structures (included as Appendix~\ref{app:wrapper}).
    \item[\texttt{mpir\_threadcomm.h}:] empty file to satisfy one of the include dependencies.
    \item[\texttt{rd\_allreduce\_driver.c}:] A driver that calls \texttt{MPIR\_Allreduce\_intra\_recursive\_doubling} and\\ \texttt{MPI\_Allreduce} and tests that the results are identical (included as Appendix~\ref{app:driver}).
    \item[\texttt{rm\_allreduce\_driver.c}:] A driver that calls \texttt{MPIR\_Allreduce\_intra\_recursive\_multiplying} \\
    and \texttt{MPI\_Allreduce} and tests that the results are identical.
    \item[\texttt{rsag\_allreduce\_driver.c}:] A driver that calls \texttt{MPIR\_Allreduce\_intra\_reduce\_scatter\_} \\
    \texttt{allgather} and \texttt{MPI\_Allreduce} and tests that the results are identical.
\end{description}

We have verified the correctness of the MPICH implementation of recursive doubling by using CIVL~\cite{CIVL} to prove the equivalence of {MPIR\_Allreduce\_intra\_recursive\_doubling} and \texttt{MPI\_Allreduce} for arbitrary inputs (up to a maximum array length of 10 for \texttt{MPI\_SUM} and \texttt{MPI\_PROD} and 5 for \texttt{MPI\_MIN} and \texttt{MPI\_MAX}) and arbitrary process counts (up to 10 and 5, respectively). In doing so, we leveraged the fact that CIVL has a builtin model of the semantics of \texttt{MPI\_Allreduce}. We needed to provide CIVL with definitions of \texttt{MPI\_Type\_size} and \texttt{MPI\_Reduce\_local} since those functions are not included in the builtin model of MPI semantics. In addition, we needed to provide a definition of the \texttt{MPIR\_Localcopy} that did not use \texttt{MPI\_Sendrecv} (alternatively, we could have removed the \texttt{const} modifier from the send buffer parameter). Finally, we needed to modify the memory allocation for a temporary buffer to use the \texttt{datatype} parameter to cast the pointer to a pointer of appropriate type (CIVL requires that all memory allocations be cast to a non-void type). We were also able to verify the MPICH implementation of reduce\_scatter-allgather following similar modifications. We do not expect to be able to use CIVL to verify the correctness of the MPICH implementation of recursive multiplying, as the latter relies on nonblocking sends and receives and CIVL cannot currently model these MPI operations.

\section*{Acknowledgements}
This material is based upon work supported by the U.S. Department of Energy, Office of Science, Office of Advanced Scientific Computing Research, under the RAPIDS Institute within the SciDAC program, under contract DE-AC02-06CH11357. We thank Hui Zhou, Ken Raffenetti, and Mike Wilkins for their insights into the MPICH implementation of allreduce.

\bibliographystyle{eptcs}
\bibliography{generic}



\newpage

\appendix

\lstset{basicstyle=\small\ttfamily,numbers=left}

\section{Example algorithm}
\label{app:allreduce}

The implementation in \texttt{MPIR\_Allreduce\_intra\_recursive\_doubling} includes a prologue and an epilogue to handle cases where the number of processes is not a power of 2. The prologue code begins at line 45. After ensuring that the number of processes participating in the reduction is a power of two, the recursive doubling algorithm begins at line 67. Finally, an epilogue at line 97 distributes the result to processes not participating in the reduction.

\begin{lstlisting}
/* Copyright (C) by Argonne National Laboratory */
#include "mpiimpl.h"
#include "mpir_threadcomm.h"

int MPIR_Allreduce_intra_recursive_doubling(const void *sendbuf,
        void *recvbuf, MPI_Aint count, MPI_Datatype datatype,
        MPI_Op op, MPIR_Comm * comm_ptr, MPIR_Errflag_t errflag)
{
    int comm_size, rank;
    int mpi_errno = MPI_SUCCESS;
    int mask, dst, is_commutative, pof2, newrank, rem, newdst;
    MPI_Aint true_extent, true_lb, extent;
    void *tmp_buf;

    MPIR_THREADCOMM_RANK_SIZE(comm_ptr, rank, comm_size);

    is_commutative = MPIR_Op_is_commutative(op);

    /* need to allocate temporary buffer to store incoming data */
    MPIR_Type_get_true_extent_impl(datatype, &true_lb, &true_extent);
    MPIR_Datatype_get_extent_macro(datatype, extent);

    MPIR_CHKLMEM_MALLOC(tmp_buf, count*(MPL_MAX(extent, true_extent)));

    /* adjust for potential negative lower bound in datatype */
    tmp_buf = (void *) ((char *) tmp_buf - true_lb);

    /* copy local data into recvbuf */
    if (sendbuf != MPI_IN_PLACE) {
        mpi_errno = MPIR_Localcopy(sendbuf, count, datatype, recvbuf, 
            count, datatype);
    }

    /* get nearest power-of-two less than or equal to comm_size */
    pof2 = MPL_pof2(comm_size);

    rem = comm_size - pof2;

    /* In the non-power-of-two case, all even-numbered
     * processes of rank < 2*rem send their data to
     * (rank+1). These even-numbered processes no longer
     * participate in the algorithm until the very end. The
     * remaining processes form a nice power-of-two. */

    if (rank < 2 * rem) {
        if (rank % 2 == 0) {    /* even */
            mpi_errno = MPIC_Send(recvbuf, count, datatype, rank + 1, 
                MPIR_ALLREDUCE_TAG, comm_ptr, errflag);

            /* temporarily set the rank to -1 so that this process 
             * does not pariticipate in recursive doubling */
            newrank = -1;
        } else {        /* odd */
            mpi_errno = MPIC_Recv(tmp_buf, count, datatype, rank - 1,
                MPIR_ALLREDUCE_TAG, comm_ptr, MPI_STATUS_IGNORE);

            /* do the reduction on received data. since the
             * ordering is right, it doesn't matter whether
             * the operation is commutative or not. */
            mpi_errno = MPIR_Reduce_local(tmp_buf, recvbuf, count, 
                datatype, op);
            /* change the rank */
            newrank = rank / 2;
        }
    } else      /* rank >= 2*rem */
        newrank = rank - rem;
    if (newrank != -1) {
        mask = 0x1;
        while (mask < pof2) {
            newdst = newrank ^ mask;
            /* find real rank of dest */
            dst = (newdst < rem) ? newdst * 2 + 1 : newdst + rem;

            /* Send the most current data, which is in recvbuf. Recv
             * into tmp_buf */
            mpi_errno = MPIC_Sendrecv(recvbuf, count, datatype, dst, 
               MPIR_ALLREDUCE_TAG, tmp_buf, count, datatype, dst,
               MPIR_ALLREDUCE_TAG, comm_ptr, MPI_STATUS_IGNORE, errflag);

            /* tmp_buf contains data received in this step.
             * recvbuf contains data accumulated so far */
            if (is_commutative || (dst < rank)) {
                /* op is commutative OR the order is already right */
                mpi_errno = MPIR_Reduce_local(tmp_buf, recvbuf, count, 
                    datatype, op);
            } else {
                /* op is noncommutative and the order is not right */
                mpi_errno = MPIR_Reduce_local(recvbuf, tmp_buf, count, 
                    datatype, op);
                /* copy result back into recvbuf */
                mpi_errno = MPIR_Localcopy(tmp_buf, count, datatype, 
                    recvbuf, count, datatype);
            }
            mask <<= 1;
        }
    }
    /* In the non-power-of-two case, all odd-numbered
     * processes of rank < 2*rem send the result to
     * (rank-1), the ranks who didn't participate above. */
    if (rank < 2 * rem) {
        if (rank % 2)   /* odd */
            mpi_errno = MPIC_Send(recvbuf, count, datatype, rank - 1, 
                MPIR_ALLREDUCE_TAG, comm_ptr, errflag);
        else    /* even */
            mpi_errno = MPIC_Recv(recvbuf, count, datatype, rank + 1,
                MPIR_ALLREDUCE_TAG, comm_ptr, MPI_STATUS_IGNORE);
    }
  fn_exit:
    return mpi_errno;
}
\end{lstlisting}

\section{Wrapper functions and macros}
\label{app:wrapper}

The following wrapper functions and macros map MPICH internal functions to standard MPI functions, enabling the MPICH allreduce implementations to be compiled as standalone functions.

\begin{lstlisting}
#include "mpi.h"
#include <stdlib.h>
#include <stdio.h>

// Macros and functions to make standalone

#define MPIC_Sendrecv(A,B,C,D,E,F,G,H,I,J,K,L,M) \
        MPI_Sendrecv(A,B,C,D,E,F,G,H,I,J,(*K),L)
#define MPIC_Send(A,B,C,D,E,F,G) MPI_Send(A,B,C,D,E,(*F))
#define MPIC_Recv(A,B,C,D,E,F,G) MPI_Recv(A,B,C,D,E,(*F),G) 
#define MPIC_Isend(A,B,C,D,E,F,G,H) MPI_Isend(A,B,C,D,E,(*F),(*G))
#define MPIC_Irecv(A,B,C,D,E,F,G) MPI_Irecv(A,B,C,D,E,(*F),(*G)) 
#define MPIC_Waitall(A,B,C) MPI_Waitall(A,(*B),C)
#define MPIR_Request MPI_Request
#define MPIR_Reduce_local MPI_Reduce_local
#define MPIR_ERR_CHECK(X)
#define MPIR_THREADCOMM_RANK_SIZE(A,B,C) \
        MPI_Comm_rank(*A, &B); MPI_Comm_size(*A, &C);
#define MPL_MIN(X, Y)  ((X) < (Y) ? (X) : (Y))
#define MPL_MAX(X, Y)  ((X) > (Y) ? (X) : (Y))
#define MPIR_CHKLMEM_MALLOC(A,B) A = malloc(B)
#define MPIR_ALLREDUCE_TAG            14
#define LOCALCOPY_TAG                 2153
#define MPIR_Comm MPI_Comm
#define MPIR_CHKLMEM_DECL(X)
#define MPIR_Op_is_commutative(X) 1
#define MPIR_CHKLMEM_FREEALL(X) 
#define MPIR_Assert(X)
#define MPIR_Errflag_t MPI_Status*
#define MPIR_Datatype_get_extent_macro(A,B) \
        MPIR_Datatype_get_extent(A,&B)
#define MPIR_Localcopy(sbuf, scount, sdatatype, rbuf, rcount, rdatatype)\
    MPI_Sendrecv(sbuf, scount, sdatatype, 0, LOCALCOPY_TAG, rbuf, \
    rcount, rdatatype, 0, LOCALCOPY_TAG, MPI_COMM_SELF, \
    MPI_STATUS_IGNORE)
#define MAX_ALIGNMENT 16

void MPIR_Type_get_true_extent_impl(MPI_Datatype dtype, MPI_Aint *lbptr, 
                                    MPI_Aint *extentptr){
   int myextent;

   *lbptr = 0;
   MPI_Type_size(dtype, &myextent);
   *extentptr = myextent;
}

void MPIR_Datatype_get_extent(MPI_Datatype dtype, MPI_Aint *extentptr){
   int myextent;

   MPI_Type_size(dtype, &myextent);
   *extentptr = myextent;
}

/* Returns int(log2(number)) */
static inline int MPL_log2(int number)
{
    int p = 0;

    while (number > 0) {
        number >>= 1;
        p++;
    }
    return p - 1;
}

static inline int MPL_pof2(int number)
{
    if (number > 0) {
        return 1 << MPL_log2(number);
    } else {
        return 0;
    }
}
\end{lstlisting}

\section{Example driver}
\label{app:driver}

The driver for testing the correctness of \texttt{MPIR\_Allreduce\_intra\_recursive\_doubling} using concrete array values is as follows.

\begin{lstlisting}
#include <mpi.h>
#include <stdio.h>
#include <string.h>
#define AR_OP MPI_SUM
#define N 10

int MPIR_Allreduce_intra_recursive_doubling(const void *, void *, 
        MPI_Aint, MPI_Datatype, MPI_Op, MPI_Comm *, MPI_Status *);

int main(int argc, char *argv[]) {
    int correct = 1, rank, size;
    double x[N], allreduce_result[N], my_allreduce_result[N];
    MPI_Comm mycomm;

    MPI_Init(&argc, &argv);
    MPI_Comm_dup(MPI_COMM_WORLD, &mycomm);
    MPI_Comm_rank(mycomm, &rank);
    MPI_Comm_size(mycomm, &size);

    for (int i = 0; i < N; i++)
        x[i] = (double)rank + i*1.0;

    MPI_Allreduce(x, allreduce_result, N, MPI_DOUBLE, AR_OP, mycomm);
    MPIR_Allreduce_intra_recursive_doubling(x, my_allreduce_result, 
        N, MPI_DOUBLE, AR_OP, &mycomm, MPI_STATUS_IGNORE);

    for (int i = 0; i < N; i++) {
        if (allreduce_result[i] != my_allreduce_result[i]) {
            correct = 0;
            break;
        }
    }

    if (correct)
        printf("Rank %d: Results match!\n", rank);
    else
        printf("Rank %d: Results do NOT match!\n", rank);

    MPI_Finalize();
    return 0;
}
\end{lstlisting}

The CIVL driver is similar, replacing \texttt{\#define N 10} on line 5 with

\lstset{basicstyle=\small\ttfamily,numbers=none}

\begin{lstlisting}
#ifndef NMAX
#define NMAX 10
#endif
$input int N;
$assume (1 <= N && N <= NMAX);
$input double x_val[N][16];
\end{lstlisting}
and the initialization of \texttt{x[i]} on line 21 with
\begin{lstlisting}
        x[i] = x_val[i][rank];
\end{lstlisting}
to treat \texttt{x[i]} as symbolic values. Finally, an assertion is added before line 28:
\begin{lstlisting}
        $assert(allreduce_result[i] == my_allreduce_result[i]);
\end{lstlisting}
\end{document}